\documentclass{emulateapj}

\def\U{$U$}

\def\V{$V$}

\def\bmi{\hbox{\it B--I\/}}
\def\bmv{\hbox{\it B--V\/}}
\def\vmi{\hbox{\it V--I\/}}
\def\feh{\hbox{\rm [Fe/H]}}

\def\min{${}^{\prime}$}
\def\sec{${}^{\prime\prime}$}

\def\afe{\hbox{\rm [$\alpha$/Fe]}}

\newcommand{\kms}{\mbox{km~s$^{-1}$}}

\def\sec{${}^{\prime\prime}$}

\def\min{${}^{\prime}$}

\newcommand{\feI}{Fe\,{\scriptsize I}}
\newcommand{\feII}{Fe\,{\scriptsize II}}
\newcommand{\feIonII}{Fe\,{\scriptsize I/II}}

\newcommand{\moog}{{\scriptsize MOOG}}
\newcommand{\marcs}{{\scriptsize MARCS}}
\newcommand{\vald}{{\scriptsize VALD}}
\newcommand{\nist}{{\scriptsize NIST}}

\def\Teff{T_{\mathrm{eff}}}
\def\feh{\hbox{[\rm{Fe/H}]}}
\def\feih{\hbox{[\rm{Fe\,{\scriptsize I}/H}]}}
\def\feiih{\hbox{[\rm{Fe\,{\scriptsize II}/H}]}}

\shorttitle{The Carina Project. V. The impact of NLTE effects on the iron content}
\shortauthors{Fabrizio et al.}

\begin{document}
\title{The Carina Project. V. The impact of NLTE effects on the iron content\altaffilmark{1}}

\author{
M.~Fabrizio\altaffilmark{2},
T.~Merle\altaffilmark{3},
F.~Th\'evenin\altaffilmark{3},
M.~Nonino\altaffilmark{4},
G.~Bono\altaffilmark{2,5},
P.~B.\ Stetson\altaffilmark{6,7},
I.~Ferraro\altaffilmark{5},
G.~Iannicola\altaffilmark{5},
M.~Monelli\altaffilmark{8,9},
A.~R.~Walker\altaffilmark{10},
R.~Buonanno\altaffilmark{2,11},
F.~Caputo\altaffilmark{5},
C.~E.~Corsi\altaffilmark{5},
M.~Dall'Ora\altaffilmark{12},
S.~Degl'Innocenti\altaffilmark{13,14},
P.~Fran\c{c}ois\altaffilmark{15},
R.~Gilmozzi\altaffilmark{16},
M.~Marconi\altaffilmark{12},
A.~Pietrinferni\altaffilmark{17},
P.G.~Prada Moroni\altaffilmark{13,14},
F.~Primas\altaffilmark{16},
L.~Pulone\altaffilmark{5},
V.~Ripepi\altaffilmark{12} and
M.~Romaniello\altaffilmark{16}
}

\altaffiltext{1}{Based on spectra retrieved from the ESO/ST-ECF Science 
Archive Facility and collected either with UVES at ESO/VLT (065.N-0378(A),
066.B-0320(A), P.I.: E.~Tolstoy) or with 
FLAMES/GIRAFFE-UVES at ESO/VLT (074.B-0415(A), 
076.B-0146(A), P.I.: E.~Tolstoy; 171.B-0520(A)(B)(C), 
180.B-0806(B), P.I.: G.~Gilmore).}
\altaffiltext{2}{Dipartimento di Fisica, Universit\`{a} di Roma Tor Vergata, via della Ricerca Scientifica 1, 00133 Rome, Italy; michele.fabrizio@roma2.infn.it}
\altaffiltext{3}{Universit\'{e} de Nice Sophia-antipolis, CNRS, Observatoire de la C\^{o}te d'Azur, Laboratoire Lagrange, BP 4229, 06304 Nice, France}
\altaffiltext{4}{INAF--Osservatorio Astronomico di Trieste, via G.B. Tiepolo 11, 40131 Trieste, Italy}
\altaffiltext{5}{INAF--Osservatorio Astronomico di Roma, via Frascati 33, Monte Porzio Catone, Rome, Italy}
\altaffiltext{6}{Dominion Astrophysical Observatory, Herzberg Institute of Astrophysics, National Research Council, 5071 West Saanich Road, Victoria, BC V9E 2E7, Canada}
\altaffiltext{7}{Visiting Astronomer, Cerro Tololo Inter-American Observatory, National Optical Astronomy Observatories, operated by AURA, Inc., under cooperative agreement with the NSF.}
\altaffiltext{8}{Instituto de Astrof\'{i}sica de Canarias, Calle Via Lactea, E38200 La Laguna, Tenerife, Spain}
\altaffiltext{9}{Departamento de Astrof\'{i}sica, Universidad de La Laguna, Tenerife, Spain}
\altaffiltext{10}{Cerro Tololo Inter-American Observatory, National Optical Astronomy Observatory, Casilla 603, La Serena, Chile}
\altaffiltext{11}{Agenzia Spaziale Italiana--Science Data Center, ASDC c/o ESRIN, via G. Galilei, 00044 Frascati, Italy}
\altaffiltext{12}{INAF--Osservatorio Astronomico di Capodimonte, via Moiariello 16, 80131 Napoli, Italy}
\altaffiltext{13}{Dipartimento di Fisica, Univiversit\`{a} di Pisa, Largo B. Pontecorvo 2, 56127 Pisa, Italy}
\altaffiltext{14}{INFN, Sez. Pisa, via E. Fermi 2, 56127 Pisa, Italy}
\altaffiltext{15}{Observatoire de Paris-Meudon, GEPI, 61 avenue de l'Observatoire, 75014 Paris, France}
\altaffiltext{16}{European Southern Observatory, Karl-Schwarzschild-Str. 2, 85748 Garching bei Munchen, Germany}
\altaffiltext{17}{INAF--Osservatorio Astronomico Collurania, via M. Maggini, 64100 Teramo, Italy}
\date{\centering drafted \today\ / Received / Accepted }

\begin{abstract}
We have performed accurate iron abundance measurements for 44 red giants (RGs)
in the Carina dwarf spheroidal (dSph) galaxy. We used archival, high-resolution
spectra ($R\sim 38\,000$) collected with UVES at ESO/VLT either in slit mode (5)
or in fiber mode (39, FLAMES/GIRAFFE-UVES).
The sample is more than a factor of four larger than any previous spectroscopic
investigation of stars in dSphs based on high-resolution ($R\ge 38\,000$)
spectra.
We did not impose the ionization equilibrium between neutral and singly-ionized
iron lines. The effective temperatures and the surface gravities were estimated
by fitting stellar isochrones in the \V, \bmv\ color-magnitude diagram. To
measure the iron abundance of individual lines we applied the LTE spectrum
synthesis fitting method using \marcs\ model atmospheres of appropriate
metallicity. For the 27 stars for which we measured both \feI\ and \feII\
abundances, we found evidence of NLTE effects between neutral and singly-ionized
iron abundances. The difference is on average $\sim 0.1$~dex, but steadily
increases when moving from the metal-rich to the metal-poor regime. Moreover,
the two metallicity distributions differ at the 97\% confidence level. Assuming
that the \feII\ abundances are minimally affected by NLTE effects, we corrected
the \feI\ stellar abundances using a linear fit between \feI\ and \feII\ stellar
abundance determinations.
We found that the Carina metallicity distribution based on the corrected \feI\
abundances (44~RGs) has a weighted mean metallicity of $\feh=-1.80$ and a
weighted standard deviation of $\sigma=0.24$~dex. The Carina metallicity
distribution based on the \feII\ abundances (27~RGs) gives similar estimates
($\feh=-1.72$, $\sigma=0.24$~dex). The current weighted mean metallicities are
slightly more metal poor when compared with similar estimates available in the
literature. Furthermore, if we restrict our analysis to stars with the most
accurate iron abundances, $\sim$ 20 \feI\ and at least three \feII\ measurements
(15 stars), we found that the range in iron abundances covered by Carina RGs
($\sim1$~dex) agrees quite well with similar estimates based on
high-resolution spectra. However, it is a factor of two/three smaller than
abundance estimates based on the near-infrared Calcium triplet. This finding
supports previous estimates based on photometric metallicity indicators.
\end{abstract}

\keywords{galaxies: dwarf --- galaxies: individual (Carina) --- galaxies: stellar content --- stars: abundances --- stars: fundamental parameters}

\maketitle

\section{Introduction}

The Carina dSph galaxy is a fundamental benchmark to constrain the formation and
evolution of dwarf galaxies and a key laboratory to improve current knowledge of
low- and intermediate-mass stars' evolutionary and pulsation properties
\citep[e.g.][] {smecker99,monelli03,bono10}. Quantitative constraints on its
stellar content require not only precise multiband photometry over the entire
body of the galaxy \citep{stetson11}, but also accurate measurements of the mean
metallicity and of the spread, if any, in metallicity.

Using the multi-fiber spectrograph ARGUS at the CTIO 4~m Blanco telescope,
\citet{smecker99} collected low-resolution ($R\sim 3\,000$) spectra for 52 Red
Giants (RGs), covering the near-infrared (NIR) Calcium triplet (CaT) region, and
found a mean metallicity of $\feh=-1.99$ and a small spread in metallicity
($\sigma_{\rm[Fe/H]}=0.25$~dex). On the other hand, \citet{koch06} collected
medium-resolution spectra ($R \sim 6\,500$) for 437 RGs with the multi-fiber
spectrograph FLAMES/GIRAFFE \citep{pasquini02} in MEDUSA mode at ESO/VLT,and
using the same diagnostic found a mean metallicity of $\feh=-1.90\pm0.01$
(metallicity scale by \citealt{zw84}) and a significant spread in metallicity
(full range $-3.0 \lesssim\feh\lesssim 0.0$). \citet{koch06} transformed the
individual equivalent widths (EWs) of the CaT lines into iron abundances using
the calibration provided by \citet{rutledge97}. In a more recent investigation,
\citet{helmi06} studied the same spectroscopic data and found a similar mean
metallicity, ($\feh=-1.7\pm0.1$), but a smaller spread in iron abundance
($-2.3\lesssim\feh\lesssim -1.3$). However, they used a different calibration
between the EW of CaT lines and metallicity \citep{tolstoy01} and different
criteria for selecting candidate Carina stars. The difference between the two
metallicity distributions appears as a steeper metal-rich tail in the latter
distribution when compared with the former one.

High-resolution spectra ($R \sim 40\,000$) are available for a sample of five
bright RG stars, collected with the slit mode of Ultraviolet and Visual Echelle
Spectrograph (UVES) at ESO/VLT \citep{dekker00}. The mean metallicity found by
\citet{shetrone03} from these spectra is $\feh=-1.64$ with
$\sigma_{\rm[Fe/H]}=0.4$~dex. The spread decreases to 0.25~dex if we remove the
most metal-rich star in the sample. Independent measurements by \citet{koch08},
using high-resolution ($R \sim 38\,000$) spectra for ten bright RGs collected
with the red arm of UVES from FLAMES/GIRAFFE fibers, give a similar mean
metallicity of $\feh=-1.69$, but a spread that is a factor of two larger
($\sigma_{\rm[Fe/H]}=0.51$~dex).

A detailed spectroscopic analysis of Carina RGs was recently
provided by \citet{lemasle12}. They used FLAMES/GIRAFFE spectra for 35 RGs
collected with grisms HR10 ($R\sim 19\,800$), HR13 ($R\sim 22\,500$) and
HR14$_{\rm new}$ ($R\sim 17\,700$). They found a mean abundance based on \feI\
lines ($\mu_{\rm[FeI/H]}=-1.63\pm0.01$, $\sigma_{\rm[FeI/H]}=0.27$~dex, weighted
mean) that agrees quite well with similar estimates available in the literature.
On the other hand, the mean abundance based on \feII\ lines
($\mu_{\rm[FeII/H]}=-1.24\pm0.01$, $\sigma_{\rm[FeII/H]}=0.31$~dex, weighted
mean) is slightly more metal-rich, but the difference is of the order of
1$\sigma$. Moreover and even more importantly, they found that their metallicity
estimates range from $\feih=-1.2$ to $\feih=-2.5$, thus supporting a
corresponding decrease in the metallicity spread. 

A more recent analysis has been provided by \citet{venn12}, using
high-resolution spectra collected either with FLAMES/GIRAFFE-UVES 
(seven RGs) or with Magellan-MIKE ($R\sim 28\,000$, blue, $\sim 22\,000$,
red, two RGs). They found, by using EWs, mean abundances based on \feI\
($\mu_{\rm[FeI/H]}=-1.79\pm0.09$, $\sigma_{\rm[FeI/H]}=0.54$~dex,
weighted mean) and on \feII\ lines ($\mu_{\rm[FeII/H]}=-1.76\pm0.09$,
$\sigma_{\rm[FeII/H]}=0.53$~dex) that agree quite well with similar
estimates available in literature.
 
Current spectroscopic investigations support the evidence that Carina may host
very metal-poor stars \citep[$\feh<-3.0$,][]{helmi06}. This evidence, if
supported by independent spectroscopic investigations, would imply that Carina
underwent a fast chemical enrichment starting in a metal-poor environment and
rapidly approaching a metal-intermediate regime ($\feh\sim-1.6$).

Photometric investigations agree with the spectroscopic measurements concerning
the mean metallicity, but suggest a small spread in metallicity
($\sigma_{\rm[Fe/H]}\sim0.25$~dex, \citealt{bono10};
$\sigma_{\rm[Fe/H]}\sim0.35$~dex, \citealt{lianou11}). 
In this work we perform a reanalysis of the Carina spectra
investigated by \citet{shetrone03}, \citet{koch08} and \citet{venn12} 
together with the analysis of a new high-resolution spectroscopic data set.
In total we have 89 spectra for 72 stars. However, 18 of them are probable
non-members, two of them are carbon stars, for six of them we have not been able
to measure the radial velocity and for two of them we have not been able to
measure the iron abundances. Overall, we provide new homogeneous iron abundance
measurements for 44 Carina RGs based on high-resolution spectra collected with
UVES (slit mode, five) and with FLAMES/GIRAFFE-UVES (multifiber mode, 39) at
ESO/VLT.

This is the largest sample of high-resolution ($R\ge 38\,000$) spectra ever
collected for a dSph galaxy. A similar spectroscopic approach was also adopted
by \citet{letarte10} for RGs in the Fornax dSph. They used FLAMES/GIRAFFE
spectra for 81 RGs collected with grisms HR10, HR13 and HR14$_{\rm new}$. A set
of spectra were also collected with the old version of the grism HR14$_{\rm
old}$ ($R\sim 28\,800$), but they were rescaled to the resolution of the spectra
collected with the grism HR14$_{\rm new}$.

\section{Observations and data reduction}
The high-resolution spectra for Carina RGs adopted in this investigation were
retrieved from the ESO Science Archive. We selected spectra from four different
ESO/VLT observing programs collected with either UVES (slit mode, nine) or
FLAMES/GIRAFFE-UVES (multifiber mode, 80).

The oldest (2000--2001) data set\footnote{ESO programs 065.N-0378(A),
066.B-0320(A), PI: Tolstoy. These are the spectra adopted by
\citet{shetrone03}.} includes nine spectra of five RGs with visual magnitudes
ranging from 17.62 to 17.92~mag. These spectra were collected with red arm of
UVES, include 37 orders and have already been analyzed by \citet{shetrone03}.
The second (2003) data set\footnote{ESO programs 171.B-0520(A)(B)(C), PI:
Gilmore. These are the spectra adopted by \citet{koch08}.} comprises individual
spectra for 33 RGs collected with the FLAMES/GIRAFFE-UVES red arm; their visual
magnitudes range from 17.32 to 18.98~mag. The orders and the wavelength coverage
of these spectra are the same as the UVES spectra. They have already been
analyzed by \citet{koch08} who found that two of them are candidate carbon stars
(stars Car54, Car55, or using Kock's IDs, LG04a\_000057 and LG04b\_000569). We
support the classification suggested by \citet{koch08}, and have {\it not\/}
included these stars in the current abundance analyses. 
The third (2005) data set\footnote{ESO programs 074.B-0415(A),
076.B-0146(A), PI: Tolstoy. These are the high-resolution spectra 
for seven RGs adopted by \citet{venn12}} and collected with
FLAMES/GIRAFFE-UVES. The visual magnitudes of the RGs range from 
17.65 to 18.00~mag.
In the fourth (2007--2008) data
set\footnote{ESO programs 180.B-0806(B), PI: Gilmore} there are 40 spectra for
40 different stars collected with FLAMES/GIRAFFE-UVES, and their visual
magnitudes range from 17.61 to 18.68~mag.

All the spectra collected with the red arm of UVES cover the wavelength range
$4780-6825$~\AA\ and the signal-to-noise ratio ($S/N$) is typically of the order
of 30 ($\lambda\sim$ 6750 \AA). The spectra collected with the
FLAMES/GIRAFFE-UVES red arm cover the same wavelengths and have $S/N$ (at
$\lambda\sim$ 6750 \AA) ranging from $\sim$15$\pm$5 (21 stars) to
$\sim$30$\pm$10 (11 stars) and $\sim$45$\pm$5 (7 stars).

We ended up with a sample of 89 high-resolution spectra for 72 stars, located
across Carina's central regions and covering the bright portion of the RG branch
(see Fig.~1). We chose to use spectra taken with the red arm of UVES (centered
at 5800~\AA) because it is minimally affected by contamination from sky lines.
The UVES spectra were reduced using IRAF\footnote{IRAF is distributed by the
National Optical Astronomy Observatory, which is operated by the Association of
Universities for Research in Astronomy, Inc., under cooperative agreement with
the National Science Foundation.}, and extracted using the IRAF task {\tt
apall}. Wavelength calibration was performed using reference lines from
\citep{murphy2007}.

\section{Radial velocity and photometric analysis}
We measured the radial velocity ($RV$) of each spectrum following the same
approach as adopted by \citet{fabrizio11}, using two dozen heavy-element lines
ranging from 6136 to 6200~\AA. We measured the $RV$ of 66 stars (81 spectra).
For six stars the measurement of the $RV$ was not possible due to the poor
quality of the spectra. Among the stars with $RV$ measurements, 46 (61 spectra)
are candidate Carina stars (with $212 < RV < 243$~\kms, \citealt{fabrizio11}).
The others are either candidate field stars (18, $RV < 100$~\kms) or candidate
carbon stars (two) and they were not included in the current analysis.

To provide accurate estimates for the stellar parameters of the spectroscopic
targets (effective temperature $\Teff$ and surface gravity $\log g$), we used
the multiband photometry discussed in \citet{bono10} and in \citet{stetson11}.
Moreover, we adopted different scaled-Solar cluster isochrones from the BaSTI
data base\footnote{Available at the URL: http://albione.oa-teramo.inaf.it/}
\citep{pietr04,pietr06}. We selected isochrones at fixed age (12~Gyr) for three
different chemical compositions ($\feh=-1.50, -1.79, -2.27$). We also adopted a
true distance modulus $DM_0=20.15$~mag and a reddening $E(B-V)=0.04$~mag
\citep{dallora03}. Finally, we performed a linear regression among the different
isochrones and derived analytical relations connecting the visual magnitude \V,
the \bmv\ color and either the surface gravity or the effective temperature. On
the basis of the above relations and of the observed magnitudes and colors, we
estimated the atmospheric parameters for each spectroscopic target. The adopted
approach and the analytical relations will be described in a future paper
(Ferraro et al. in preparation).

The top panel of Fig.~1 shows the position of the spectroscopic targets in the
\V, \bmv\ Color-Magnitude Diagram (CMD) together with the entire photometric
catalog (grey dots, \citealt{bono10}). The black dots mark the 46 candidate
Carina stars, while the tiny orange dots display the 20 candidate field stars
and the two large orange dots show the carbon stars. The blue squares and the
red stars show the high-resolution sample of \citet{koch08} and
\citet{shetrone03}, respectively. The solid lines display the adopted
isochrones.
Taking into account current uncertainties in the Carina true distance modulus,
the reddening, the mean metallicity \citep{dallora03,pietrynski09,
bono10,lemasle12} and the spread in age \citep{stetson11}, we performed a series
of simulations and found that the typical uncertainties in temperature and
gravity are $\epsilon_{\Teff} \sim 70$~K and $\epsilon_{\log{g}} \sim 0.2$~dex
(see Table~1). The bottom panel of Fig.~1 shows the same isochrones as the top
panel, but in the $\log g$ vs $\Teff$ plane together with the target stars. The
error bars plotted in the bottom right corner represent the aforementioned
uncertainties.
\section{Spectroscopic analysis}
The iron abundance analysis was performed following the classical
spectrum-synthesis method for both \feI\ and \feII\ lines, but with one
difference:  we did not impose LTE ionization equilibrium between the \feI\ and
\feII\ lines (e.g. \citealt{ki03}). This means that we trust the surface gravity
determined from the optical photometry and cluster isochrones. For the highest
$S/N$ spectra collected with UVES, we selected \feI\ and \feII\ lines from the
\vald\footnote{Available at the URL: http://vald.astro.univie.ac.at} data base
\citep{kupka00}.
We ended up with a list of 123 \feI\ and 18 \feII\ lines in the wavelength range
covered by our spectra. Among them, 45 \feI\ and 5 \feII\ lines are located in
overlapping orders.

By using the photometric estimates of both $\Teff$ and $\log g$, and the Carina
mean metallicity $\feh=-1.70$ \citep{koch08}, we interpolated the
\marcs\footnote{Available at the URL: http://marcs.astro.uu.se} model
atmospheres \citep{gustafsson08} with a modified version of the
\citet{masseron06} code\footnote{Available at the URL:
http://marcs.astro.uu.se/software.php} for each Carina star. The individual
models assumed spherical geometry, an $\alpha$-enhanced ($\afe=+0.4$) chemical
mixture, a mass value of 1$M_\odot$\footnote{Note that for stellar masses
ranging from 0.8 to 1.2$M_\odot$, the change in $\log g$, at fixed $\Teff$, is
smaller than 0.1~dex} and a constant microturbulence velocity ($\xi=2$~\kms).
The value adopted for the microturbulence velocity follows the estimates
provided by \citet{shetrone03} for Carina red giants using metal-poor stellar
model atmospheres. Similar estimates for the $\xi$ values were provided by
\citet{thevenin98} for RG stars with similar atmospheric parameters in the
globular $\omega$ Cen. The Carina targets span a modest range in atmospheric
parameters, therefore we decided to adopt the same $\xi$ value for the entire
sample.
The synthetic spectra were computed using the 1D, plane-parallel, LTE, radiative
transfer code \moog\footnote{Available at the URL:
http://www.as.utexas.edu/$\sim$chris/moog.html} (2009 version,
\citealt{sneden73}). Note that plane-parallel radiative transfer for giant stars
is more accurate when using spherical model atmospheres \citep{heiter06}.

We chose the Solar chemical composition from \citet{grevesse07}, i.e., a Solar
iron abundance of $A_{{\rm Fe}}^{\odot} = 7.45$\footnote{$A_{{\rm Fe}} = \log
(N_{{\rm Fe}}/N_{{\rm H}}) + 12$ where $N_{\rm Fe}$ and $N_{\rm H}$ are the
number densities of Fe and H}. Oscillator strengths $f$ for \feI\ and \feII\
lines are from \vald. When available, the parameters for collisional damping
with neutral hydrogen are based on quantum theory
\citep{anstee95,barklem97,barklem98}; otherwise the Uns\"old classical recipe
\citep{unsoeld55} was used with an enhanced damping factor of $F_H = 1.5$. We
note that calculations based on quantum theory are available for all the
selected \feII\ lines. The strategy adopted for the iron abundance analysis was
dictated by the quality of the observations: high-resolution ($R\sim 38\,000$)
and a reasonable signal-to-noise ratio ($15 \lesssim S/N \lesssim 45$). We
decided to use an eye fitting procedure rather than an automatic EW fitting
procedure, since most of the weak lines are either noisy or blended.

The selected \feI\ and \feII\ lines were normalized using a linear fit of the
local continuum identified by eye inspection. Using the \moog\ code, we
synthesized spectra with different metallicities. The synthetic spectra were
convolved with a Gaussian broadening function to reproduce the instrumental
resolution. We neglected the effect of stellar rotation. The fit of the lines
(45~\feI\ + 5~\feII) that appear in two overlapping orders were treated
independently and the iron abundance for these lines was estimated as a weighted
mean. After a preliminary analysis of the stars with multiple spectra, we
decided to neglect the spectra with lower $S/N$ and we only analyzed the best
available spectrum for each star (46 spectra).

It is worth mentioning that in the abundance analysis we did not include lines
with $\lambda < 4920$~\AA\, to overcome subtle uncertainties with the coherent
scattering by hydrogen in the continuum that was not included in \moog\
\citep{sobeck11}. To constrain quantitatively the possible error in abundance in
this wavelength range we computed the abundances of the three \feI\ lines at
4924, 4950 and 4973~\AA. We used a \marcs\ model atmosphere assuming
$\Teff=4500$~K, $\log g=1$~ dex and $\feh=-1.5$ with the code {\scriptsize
TURBOSPECTRUM}\footnote{Available at the URL:
http://www.graal.univ-montp2.fr/hosted/plez}\citep{alvarez98} which accounts for
the coherent scattering by hydrogen in the continuum. We found no significant
difference with the abundance measurement based on the \moog\ code, and indeed
the differences in the cores of the lines represent less than five percent of
the line equivalent widths.

We performed a first preliminary mean iron abundance analysis for each star.
Subsequently, we computed a new set of model atmospheres using the same values
of $\Teff$ and $\log~g$, but using the new individual iron abundances. We also
decided to use scaled-Solar ($\afe = 0.0$) \marcs\ models, since detailed
abundance measurements of RGs in nearby dwarf galaxies \citep{shetrone03,koch08}
indicates that they show only a mild enhancement in $\alpha$-elements. For stars
with $\feh < -2.0$, we used $\alpha$-enhanced models because the very
metal-poor, scaled-Solar \marcs\ models are not available. In this new iron
abundance analysis, we double checked all the iron measurements and rejected
strong, noisy, blended or too weak \feI\ and \feII\ lines from the analysis.
The adopted line list with wavelengths for both neutral and ionized iron lines
is presented in Table~2 together with ion identification, excitation potential
(EP) and $\log gf$ values. We ended up with 83 \feI\ and 11 \feII\ lines.
Among them, 25 \feI\ and 4 \feII\ lines are located in overlapping orders. We
performed several experiments with synthetic spectra and we found that the
accuracy of the line fits ranges from 0.1 to 0.15~dex.

The atmospheric uncertainty in the line-by-line abundance measurements was
estimated using the formula based on the EW uncertainty suggested by
\citet{cayrel88}. In particular, we adopted a mean equivalent width
$\rm\overline{EW}$ (typical values of the selected lines are $\sim 50$~m\AA\ for
\feI\ and $\sim 30$~m\AA\ for \feII) and the $S/N$ of each individual spectral
order (estimated using the IRAF task {\tt splot}). The final iron abundance for
each star was computed as the weighted mean of the line-by-line abundances.
Individual weights were chosen as the inverse of the atmospheric uncertanties to
limit the influence of the outliers.

For each star, in Table~2 are also listed the line-by-line EWs. The
EWs were measured using the IRAF task {\tt splot}, assuming a single
Gaussian line profile, since the selected lines are not blended nor
asymmetric. Typical uncertainty on the widths is of the order of 5\%,
and it is caused by the estimate of the continuum. It is worth mentioning
that the EWs listed in Table~2 are only given for completeness of
the current spectroscopic analysis. The iron abundances we provide are 
only based on the spectrum synthesis fitting method. The differences 
between our EWs and those measured by \citet{shetrone03} and \citet{koch08}
are minimal (weighted mean, $\rm{EW_{our}-EW_{author}} \sim -5.4$ and
$\sim 1.7$~m\AA, respectively). The difference with EWs provided by
\citet{lemasle12} is larger (eleven stars in common, $\sim -22$~m\AA).
The comparison between the EWs provided by \citet{lemasle12} and those
by \citet{koch08} gives a similar difference (one star in common, $\sim
-27$~m\AA). The difference with EWs provided by
\citet{venn12} is quite smaller (seven stars in common, $\sim -13$~m\AA), 
and compared with EWs provided by \citet{koch08} gives a similar difference 
(three stars in common, $\sim -18$~m\AA).

\section{Results and discussion}
\subsection{Detailed \feI\ and \feII\ stellar abundances}
We measured \feih\ for 44 stars and \feiih\ for 27 stars out of the 46 stars in
our sample. On average, the former are based on more than twelve measurements
(five stars have a single measurement), and the latter are based on more than
three measurements (eight stars have a single measurement). The iron abundance
was not measured in two metal-poor stars (Car17, Car26) 
because the lines available were noisy, blended or misidentified.
The individual \feih\ and \feiih\ abundance measurements and their atmospheric
errors for the entire sample are listed in Table~2. The mean \feih\ and \feiih\
abundances for the entire sample, with their weighted standard deviations are
given in columns 10 and 12 of Table~1\footnote{Columns 10 and 12 give the
atmospheric uncertainty for the stars with only one line measurement.}.
Columns 11 and 13 list the number of \feI\ and \feII\ lines used to estimate the
individual weighted mean abundances.

We carried out a more detailed analysis to constrain the impact that the adopted
atmospheric parameters have on individual abundance measurements. To address
this classical issue, we computed synthetic spectra with a series of model
atmospheres at fixed chemical composition, changing the effective temperature by
100~K, the surface gravity by 0.3~dex and the microturbulence velocity by
0.5~\kms. Note that we are using generous estimates for the uncertainties
affecting the atmospheric parameters (see \S 3). In particular, we selected the
star Car23, since its effective temperature ($\sim 4400$~K) and surface gravity
($\sim 0.80$~dex) can be considered as average values for the entire sample. The
results listed in Table~3 indicate that a change of 100~K in effective
temperature causes a change in the mean \feI\ and \feII\ abundances of the order
of 0.1 dex. On the other hand, a change of 0.3~dex in $\log g$ has a minimal
impact on the \feI\ abundance and a difference of the order of 0.15~dex for the
\feII\ abundance. The differences in abundance caused by a change of 0.5~\kms\
in microturbulence velocity are of the order of 0.05 and 0.07~dex for the \feI\
and \feII\ abundances, respectively. These findings support the analysis
performed by \citet{koch08} and indicate that plausible changes in the
atmospheric parameters introduce changes in the inferred iron abundances smaller
than 0.15~dex. To account for the uncertainties affecting the atmospheric
parameters, the errors on the weighted mean metallicities listed in column 10
and 12 of Table~1 were estimated by summing in quadrature the weighted standard
deviation with the three typical errors stated above.

The iron abundances of the 27 stars for which we have both \feih\ and \feiih\
measurements (Table~1) indicate that the \feih\ abundances are systematically
lower than \feiih. The mean difference ranges from $\sim$0.1 to 0.2~dex (Car12,
Car3). Five stars show an opposite trend, but their \feII\ measurements are
based on one or two lines, and two out of the five have \feI\ measurements based
on two lines.

The detailed iron abundance analysis for five Carina stars, covering the entire
range of apparent magnitudes, is shown in Fig.~2. To validate the accuracy of
our $\Teff$ determinations, we plotted for each star the abundance of individual
lines as a function of the Excitation Potential (EP) of the lines. We performed
linear fits and found that the slopes are within $\pm0.06$~dex/eV (see labeled
values in Fig.~2), except for a few stars with a limited number of \feI\
measurements. We tested the \feI\ lines (left panels), since they are based on a
larger number of measurements. The data plotted in Fig.~2 indicate that weighted
standard deviations for \feI\ and \feII\ lines are on average $0.12$ and
$0.18$~dex, respectively. We also checked the impact of a change in
microturbulence velocity on star Car14 by assuming $\xi=2.5$~\kms. The slope
changed by less than 0.02~dex/eV, since the saturated \feI\ lines were not taken
into account in the current analysis.

\subsection{Comparison with previous investigations}

To validate our abundance analysis, we performed a detailed comparison with
similar measurements available in the literature. We defined the iron abundance
difference $\Delta\feh=\feh_{\mathrm{our}} -\feh_{\mathrm{author}}$ after
shifting the $\feh_{\mathrm{author}}$ to the Solar iron abundance that we used
($A_{\rm Fe}^\odot=7.45$, \citealt{grevesse07}). The data plotted in Fig.~3 show
$\Delta\feih$ (left) and $\Delta\feiih$ (right) versus our iron abundance. The
vertical error bars are the sum in quadrature of our atmospheric dispersion and
the errors given by the different authors, while the dotted lines display the
1~$\sigma$ difference.

We have five stars in common with \citet{shetrone03} (Car2, Car3, Car4, Car10,
Car12). Their Solar iron abundance reference is 7.52 from \citet{grevesse98}.
The data plotted in the panels a) of Fig.~3 show that our iron abundances are
systematically more metal-poor than those by \citet{shetrone03}, especially for
\feI. The weighted mean and weighted standard deviation of the difference are
$\mu_{\Delta \mathrm{[FeI/H]}}=-0.25\pm0.07$, $\sigma_{\Delta
\mathrm{[FeI/H]}}=0.17$ and $\mu_{\Delta \mathrm{[FeII/H]}}=-0.17\pm0.11$,
$\sigma_{\Delta \mathrm{[FeII/H]}}=0.14$~dex. Star Car3\footnote{The reader
interested in more detailed analysis of the photometric and spectroscopic
properties of Car3 is referred to the Appendix.}, the object with the largest
discrepancy, suggests that the iron abundance from these authors is an
overestimate. This hypothesis is supported by \citet{koch06} who found
$\feh=-1.93$ after accounting for the difference in the zero-point. This agrees
better with our abundances ($\feih=-2.14$, $\feiih=-1.97$) than with the
Shetrone's estimates ($\feih=-1.58$, $\feiih=-1.56$, using our iron Solar
abundance). Neglecting this star, our abundances agree quite well with the
\citet{shetrone03} measurements, and indeed the differences decrease to
$\mu_{\Delta \mathrm{[FeI/H]}}=-0.17\pm0.08$, $\sigma_{\Delta
\mathrm{[FeI/H]}}=0.07$ and $\mu_{\Delta \mathrm{[FeII/H]}}=-0.09\pm0.12$,
$\sigma_{\Delta \mathrm{[FeII/H]}}=0.04$~dex. We adopted the same \moog\
radiative transfer code and the same \marcs\ model atmospheres (values of
$\Teff$ and $\log g$ similar), but we used more recent versions (2009 for \moog\
and 2008 for \marcs).
The microturbulence velocities adopted by \citet{shetrone03} range from 1.9 to
2.2~\kms\ and are quite similar to the value we adopted. We also performed a
series of tests to constrain the difference between the results of
\citet{shetrone03} and the current abundances. 
The differences of EWs measured for this work (see Table~2) and
those published by \citet[][see their Table~4]{shetrone03} is minimal
(a few m\AA, see \S4).
The oscillator strengths
they adopted come from the papers of the Lick-Texas group
(\citealt{fulbright00}, and references therein) and from the \nist on-line
Atomic Spectra Database\footnote{Available at the URL:
http://physics.nist.gov/cgi-bin/AtData/main\_asd}. We have already mentioned
that our atomic data come from the \vald\ database, and it is very difficult to
critically evaluate the accuracy of atomic data coming from different databases.
This is the reason why we give a very low statistical weight to the stars in our
sample with iron abundances based on only one or two lines. This evidence
indicates that the discrepancies might result from the different approaches
adopted to estimate the abundances. Their abundances were estimated using the EW
fitting method and forcing ionization equilibrium between \feI\ and \feII,
whereas we adopted spectrum synthesis fitting method and did not force
ionization equilibrium. It is also worth noting that the number of \feI\ and
\feII\ lines adopted by \citet{shetrone03} is three times larger than our
selection.

We also have ten stars in common with \citet{koch08}. In their spectroscopic
analysis (see also \citealt{koch06}), these authors adopted the metallicity
scale of \citet{cg97}. Therefore, we assumed that they adopted a Solar iron
abundance of 7.52. The comparison, after the shift in the zero-point, is shown
in panels b) of Fig.~3. We found that we agree, within 1$\sigma$, with
these authors for the \feII\ abundance with $\mu_{\Delta
\mathrm{[FeII/H]}}=-0.21\pm 0.08$ and $\sigma_{\Delta
\mathrm{[FeII/H]}}=0.27$~dex. For \feI\ the discrepancy is larger, with
$\mu_{\Delta \mathrm{[FeI/H]}}=-0.34\pm 0.07$ and $\sigma_{\Delta
\mathrm{[FeI/H]}}=0.26$~dex.
In this case the difference in the abundances might be due to a difference
either in the approach (EW versus spectrum-synthesis fitting methods) or in the
input physics (or both). As a matter of fact, \citet{koch08} adopted their
oscillator strengths from the standard RG star Arcturus. Moreover, they used
\citet{castelli03} model atmospheres, while we used \marcs\ models and their
surface gravities are on average 0.3--0.5~dex higher than ours. The difference
in gravity is the consequence of the different approach in estimating the
gravity (forcing the balance between \feI\ and \feII\ versus photometric
gravities).

We compared the current \feI\ and \feII\ abundances with similar abundances
recently provided by \citet{lemasle12} from FLAMES/GIRAFFE spectra with a
spectral resolution that is on average a factor of two smaller than ours. The
data plotted in the panels c) of Fig.~3 show that the differences, after
correcting for their reference Solar iron abundance \citep{grevesse98}, in the
\feI\ abundances ($\mu_{\Delta\mathrm{[FeI/H]}}=-0.27\pm0.09$~dex, weighted
mean) and in the dispersion ($\sigma_{\Delta\mathrm{[FeI/H]}}=0.16$) for the
eleven stars in common are quite similar to the results based on the other
high-resolution estimates. The difference in the \feII\ abundances is based on
only four objects and attains similar values, namely
$\mu_{\Delta\mathrm{[FeII/H]}}=-0.61\pm0.27$,
$\sigma_{\Delta\mathrm{[FeII/H]}}=0.25$~dex. Note that our \feiih\ abundance for
the star with the largest discrepancy is based on only three \feII\ lines.

Finally, we compared our \feI\ and \feII\ abundances with similar
abundances provided by \citet{venn12} using EWs of FLAMES/GIRAFFE-UVES 
spectra. The data plotted in the panels d) of Fig.~3 show that 
the \feI\ abundances for the seven stars in common, after correcting 
for their reference Solar iron abundance \citep{asplund09}, 
($\mu_{\Delta\mathrm{[FeI/H]}}=-0.37\pm0.11$~dex, weighted mean) 
and their dispersion ($\sigma_{\Delta\mathrm{[FeI/H]}}=0.19$) 
agree quite well with current estimates. 
The difference in the \feII\ abundances is based on six objects 
and attains similar values, namely
$\mu_{\Delta\mathrm{[FeII/H]}}=-0.26\pm0.13$,
$\sigma_{\Delta\mathrm{[FeII/H]}}=0.11$~dex.

We also compared our iron abundances to similar estimates by \citet{koch06}.
These authors performed detailed measurements of CaT EW for 437 RGs and
transformed the reduced EWs into iron abundances, using the metallicity scales
from \citet[][ZW84]{zw84} and \citet[][CG97]{cg97}. We compared their \feI\
abundance estimates with our measurements of \feI\ abundances. We have 25 stars
with at least two iron line measurements in common with this sample and we found
that the differences in the ZW84 and the CG97 metallicity scales are:
$\mu_{{\Delta \mathrm{[Fe/H]}}}=-0.14\pm 0.04$ ($\sigma_{{\Delta
\mathrm{[Fe/H]}}}=0.15$) and $\mu_{{\Delta\mathrm{[Fe/H]}}}=-0.33\pm 0.04$
($\sigma_{{\Delta\mathrm{[Fe/H]}}}=0.18$)~dex (see Fig.~4). Our iron abundances
are slightly more metal-poor, but within 1~$\sigma$ and in better agreement with
the ZW84 than with the CG97 metallicity scale. The difference of 0.2~dex between
the ZW84 and the CG97 scale is well known \citep{cg97,ki03}.

\subsection{Carina metallicity distribution}

The data plotted in the top panels of Fig.~5 show the weighted metallicity
distributions of our sample based on \feI\ (left) and on \feII\ (right)
measurements. To constrain the metallicity distribution, we assigned to each
star a Gaussian kernel \citep{dicecco10} with a $\sigma$ equal to the standard
deviation of the iron abundance measurement. The solid lines were computed by
summing the individual Gaussian over the entire data set (see panels a and b in
Fig.~5). We found that both the median ($-1.84$~dex) and the weighted
mean\footnote{ The weighted mean metallicities are estimated by weighting the
iron abundance of each star with the quadratic inverse of the individual
uncertainty.} ($\mu_{\mathrm{[FeI/H]}}=-1.90\pm0.02$~dex) of \feI\ abundances
(44 stars) attain similar values. The same outcome applies to the \feII\
measurements (27 stars) for which we determined
$\mu_{\mathrm{[FeII/H]}}=-1.72\pm0.04$~dex. The weighted standard deviations
attain small values ($\sim0.25$~dex), but the metallicity distributions cover at
least one dex. We also estimated the metallicity distribution by giving equal
weight to all stars, and the difference with the weighted mean is minimal ($<
0.04$~dex).

To constrain the dependence of these results on the accuracy of the abundance
analysis, we recomputed the metallicity distribution using only stars with more
than twelve \feI\ and two \feII\ line measurements (see panels c and d in
Fig.~5). The weighted means and the weighted standard deviations reach values
similar to those from the entire sample, but the range in metallicity decreases
by at least 0.2~dex. The difference is mainly in the metal-poor tail, and
indicates that high-quality spectra are required to constrain the actual
dispersion in iron abundance of Carina RGs. To further constrain the impact of
the abundance precision on the metallicity distribution we selected the stars
with at least three \feII\ line measurements. The \feI\ metallicity distribution
of the same stars is typically based on $\sim20$ \feI\ lines. We ended up with a
sample of 15 stars, and the weighted mean for the \feI\ metallicity distribution
is slightly more metal-rich $\mu_{{\mathrm{[FeI/H]}}}=-1.77\pm0.04$~dex, while
the $\sigma_{{\mathrm{[FeI/H]}}}$ decreases to 0.19~dex. The same outcome
applies to the \feII\ metallicity distributions. Interestingly enough, the \feI\
and \feII\ metallicity distributions (see panels e and f in Fig.~5) cover a
range of $\sim1$~dex. The use of the best data indicates that the spread in
metallicity of Carina RGs is smaller than previous estimates. Note that this
finding appears to be minimally affected by selection biases, since the
metal-poor stars ($\feih\le-2.20$) not included in the metallicity distribution
of panel e) cover the same range in luminosity as the selected
stars\footnote{The star Car40 is slightly fainter 
than the selected stars (\V=18.56), but the iron abundance 
is based on a single line.}. 
However, these findings call for new abundance analyses to constrain the 
real extent of the metal-poor tail of Carina.

By taking into account the 27 stars for which we determined both \feI\ and
\feII\ abundances, we found that the weighted mean based on \feI\ lines is
0.12~dex lower than the weighted mean based on \feII\ lines. The difference
brings forward the possible occurrence of NLTE effects between \feI\ and \feII\
abundances. Our experiment appears appropriate to constrain this effect, since
we relaxed the constraint on the ionization equilibrium between \feI\ and \feII\
lines. This is an interesting finding, since detailed calculations and the
comparison between theory and observations
\citep{thevenin99,thevenin01,mashonkina11} show that the NLTE effects strongly
affect the less abundant species (i.e., \feI) which is over-ionized in RGs, in
particular in metal-poor stars. This effect explain why the LTE ionization
equilibrium between neutral and singly-ionized iron is destroyed and why we do
not force \feih\ to be equal to \feiih\ by decreasing the surface gravity.
Note that for the five stars in common with \citet{shetrone03} we have on
average larger $\log g$ values by 0.2~dex, but they decreased their photometric
estimates of $\log g$ by $-0.29$~dex. The decrease in the $\log g$ values
artificially forces the ionization equilibrium and causes an increase in their
\feI\ abundance determination. This approach partially overcome the NLTE effects
by changing the surface gravity. The more abundant species (i.e., \feII) are
supposed to be free from NLTE effects, at least on the ionization equilibrium.

\subsection{The impact of NLTE effects}
To further constrain this effect, we directly compared \feI\ and \feII\
abundances. Data plotted in panels a), b) and c) of Fig.~6 indicate that the
difference appear to increase when moving from metal-rich to metal-poor stars.
We performed a linear fit (panel a, red dashed line, the zero-point $\alpha$ and
the slope $\beta$ are labeled) over the 27 stars for which we have both \feI\
and \feII\ measurements. We found that the two metallicity distributions differ
at the $\sim$97\% confidence level. To further constrain this effect, we
performed the same analysis, but using stars with abundances based on at least
two \feII\ lines (panel b). The sample decreased to 19 stars, but the two
distributions still differ at the $\sim$87\% confidence level. The outcome is
the same if we use stars with abundances based on at least three \feII\ lines
(panel c), 15 stars, $\sim$75\% confidence level.

On the basis of this empirical evidence, we corrected the \feI\ abundances to
account for the NLTE effects by using the linear fit between \feI\ and \feII\
abundances plotted as a red dashed line in panel a) of Fig.~6.
The new \feI\ metallicity distributions accounting for the NLTE effects are
plotted as red dashed lines in panels a), c) and e) of Fig.~5. The weighted mean
\feI\ metallicity increases by $\sim$0.1~dex, while the weighted standard
deviations attain similar values. The use of linear fits based on more accurate
\feII\ abundances for either 19 (panel b) or 15 (panel c) stars gives similar
changes in the metallicity distributions\footnote{The weighted mean
metallicities and the weighted standard deviations based on the linear fit
plotted in panel c) of Fig.~6 are:
$\mu(\mathrm{NLTE})=-1.75\pm0.02$, $\sigma(\mathrm{NLTE})=0.23$~dex [44 stars],\\
$\mu(\mathrm{NLTE})=-1.71\pm0.03$, $\sigma(\mathrm{NLTE})=0.24$~dex [20 stars],\\
$\mu(\mathrm{NLTE})=-1.64\pm0.04$, $\sigma(\mathrm{NLTE})=0.17$~dex [15 stars].}. 
These results indicate that the weighted mean metallicity of the Carina RGs,
corrected for NLTE effects, ranges from $\mu(\mathrm{NLTE})=-1.68$ (15 stars) to
$-1.80$ (44 stars), while the weighted standard deviation ranges from 0.18 to
0.24~dex. The extreme range in iron abundance covered by the stars with the most
accurate measurements is $\sim$1~dex.

In order to constrain the NLTE effects we performed NLTE computations for two
metal abundances ($\feh=-2.0, -1.5$) using the \feIonII\ atom model provided by
\citet{collet05}. Note that to constrain the slope of the NLTE effects between
\feI\ and \feII\ lines, the two \marcs\ model atmospheres used in NLTE
computations have the same effective temperature (4500~K) and the same surface
gravity (1.0~dex). The calculations were performed following the approach
adopted by \citet{thevenin99} and by \citet{collet05}. Moreover, we did not use
inelastic collisions for supergiants as suggested by \citet{merle11}, so our
errors on NLTE abundances have to be considered as upper limits. We found
significant NLTE effects in both \feI\ and \feII\ lines. Moreover, we found a
significant relative difference between \feih\ and \feiih\ theoretical
abundances. A glance at the data plotted in Fig.~7 displays that
the predicted slope agrees quite well with the observed one. In particular, the
NLTE computations show stronger effects on \feI\ than on \feII\ abundances
($\sim 0.1$~dex). However, more detailed calculations accounting for changes in
effective temperature and gravity are required before we can reach firm
conclusions concerning the impact of NLTE effects on \feI\ and \feII\
abundances.
The current findings do not demonstrate but further support the results 
obtained by \citet{bergemann11} concerning the occurrence of NLTE 
effects in metal-poor RG atmospheres. Finally, we note that our preliminary 
calculations also show non-negligeable NLTE effects on the \feII\ lines. 
This finding, once confirmed by more detailed and independent calculations, 
implies that the true mean \feiih\ abundance of Carina stars should be 
increased by at least 0.1~dex.

\section{Final remarks}
The current findings soundly support previous estimates of Carina's mean
metallicity based on the difference in color between the Red Clump (RC) stars
and the horizontal-branch (HB) at mean color of RR~Lyrae $\feh(\Delta
(B-I)^{RC}_{HB})=-1.70\pm0.19$ \citep{bono10}. The same outcome applies to the
mean metallicity estimated using stellar isochrones $\feh \sim -1.79$ (BaSTI
data base, \citealt{stetson11}), $\feh \sim-1.79\pm0.35$ \citep{lianou11}.
High-resolution spectra for 15 stars support, together with similar results
available in the literature \citep{lemasle12}, a significant decrease in the
range in iron abundances when compared with similar measurements based on CaT
($\sim1$ vs 2--3~dex). The same conclusion applies to the comparison with the
metallicity distribution predicted by chemical evolution models based on star
formation histories available in the literature \citep{lanfranchi06}.

Our sample is more than a factor of four larger than any previous spectroscopic
investigation based on high-resolution spectra. However, the current data do not
allow us to determine whether the spread is either atmospheric, i.e., caused by
a difference in the mean metallicity between the old and the intermediate-age
population, or by measurement errors. To assess whether the different stellar
populations are also characterized by different mean metallicities, new spectra
with high $S/N$ down to RC (intermediate-mass) and to red HB (low-mass) stars
are required.

We determined the mean iron abundance for a sample of 44 Carina RG stars. We
corrected for NLTE effects on the ionization equilibrium and we found
$\feh=-1.80$ and $\sigma=0.24$~dex. If this estimate of Carina's mean
metallicity is supported by future, the position of this dSph in the
Metallicity-Luminosity diagram will be between one and two $\sigma$ more
metal-rich than expected according to the empirical relation followed by dSph,
dE/dS0 and giant early type galaxies \citep{mateo08,chilingarian11}. If this
turns out to be the case, it might open new issues concerning the interplay
between chemical evolution and enrichment, star formation history and stellar
evolution in gas-poor stellar systems \citep{walker09,revaz12}.


\acknowledgments
It is a real pleasure to thank an anonymous referee for his/her pertinent
suggestions and criticisms that helped us to improve the content and the
readability of the paper. MF thanks the OCA for support as a science visitor. TM
is granted by the OCA and R\'egion PACA and supported by Thal\`es Alenia Space.
MM is supported by the Spanish Education and Science Ministry MEC
(AYA2007-67913), MEC (AYA2010-16717). It is pleasure to thank Mathieu van
Swaelmen for computing for us the synthetic iron lines with {\scriptsize
TURBOSPECTRUM}.

\appendix

The spectroscopic target Car3 that we have in common with the sample collected
by \citet[][Car3]{shetrone03} deserves a more detailed discussion. According to
\citet{bono10} and to \citet{stetson11} this star has the following photometry:
$U=20.700\pm0.033$, $B=19.172\pm0.001$, $V=17.675\pm0.003$ and
$I=16.102\pm0.003$~mag. The position of the star in the \U, \bmi\ CMD suggests
that it might not be a Carina member, but a foreground field star. The \V, \bmi\
CMD and the \bmv, \vmi\ color-color diagram support the same evidence, but not
as strongly. The star has five fainter companions within 5\sec, all of them with
$\Delta V > 4.8$~mag, and a distance $> 2.4$\sec. The star is located 2.5\min\
N-NE of the galaxy center. The neighborhood stars are sufficiently faint and
distant that the luminosity contamination on Car3, with a typical seeing of
1\sec, is at most of the order of one part in $10^4$. We also checked individual
photometric measurements and we found marginal evidence of variability.

The radial velocity of Car3 is $231.0\pm0.8$~\kms, i.e. very similar to the mean
radial velocity of candidate Carina stars
($RV_{\mathrm{mean}}=220.9\pm0.1$~\kms, \citealt{fabrizio11}). To constrain
whether Car3 is a giant or a dwarf we also forced the balance between \feI\ and
\feII\ lines and we found that the new estimate of the surface gravity is within
0.1~dex from the estimate based on the photometry ($\log g=0.45\pm0.19$~dex).
This finding further support the evidence that Car3 is a truly red giant.

According to \citet{shetrone03} Car3 is underabundant in $\alpha$-elements, and
in particular in Ca. To further constrain this evidence we selected three Carina
RGs 
(Car4, Car37, Car50) with similar surface gravities ($\log g=0.53\ \mathrm{to}\
0.67$~dex) and effective temperatures ($\Teff = 4220\ \mathrm{to}\ 4290$~K). We
performed a detailed Ca abundance analysis by using two different multiplets,
namely $\lambda=6102,\: 6162,\: 6162$~\AA\ and $\lambda=6166,\: 6169.0,\:
6169.5$~\AA. We also accounted for NLTE effects \citep{merle11} and we found
that Car3 is on average a factor of two more Ca poor that the other three Carina
RGs. We also visually inspect the entire spectrum and we did not find evidence
of molecular bands, thus supporting the absence of this star in the list of
Carina carbon stars provided by \citet{mould82} and by \citet{koch08}.
Interestingly enough, we found that Car3 shows a strong $H_\alpha$ line with
both a blue and a red P-Cygni profile. Thus, suggesting that this star might
have a complex atmosphere, probably affected by chromospheric activity. Car3 is
an object that deserves further photometric and spectroscopic investigations.



\begin{deluxetable}{llllccclccrcr}
\end{deluxetable}

\clearpage
\begin{figure}[!ht]
\begin{center}
\includegraphics[width=0.99\textwidth]{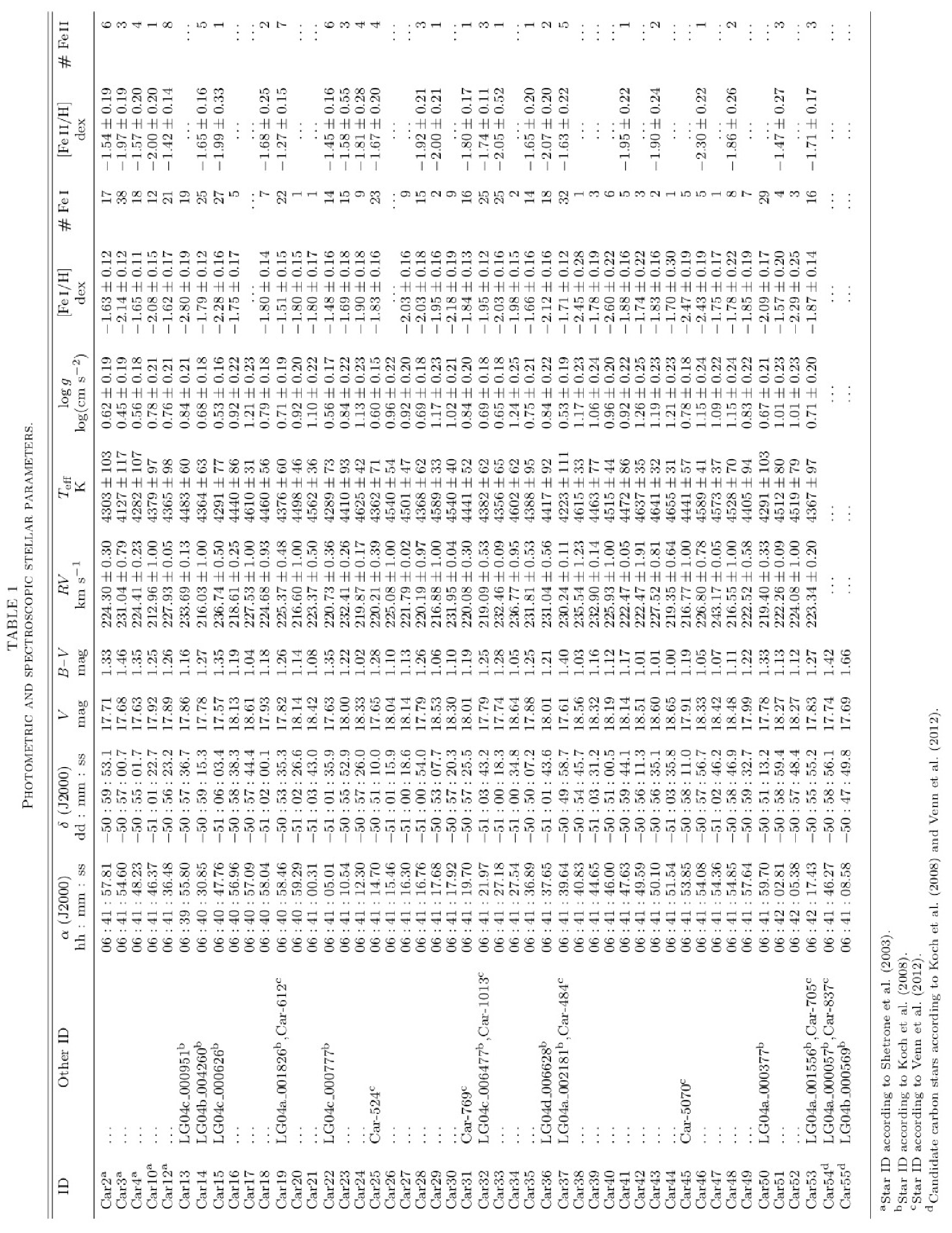}
\end{center}
\end{figure}
\clearpage

\begin{deluxetable}{llllccccccccccccccccccc}
\tabletypesize{\scriptsize}
\tablewidth{0pt}
\tablecaption{Atomic Data and Individual Iron Abundances\tablenotemark{*}}
\label{tab_lines}
\tablehead{
\colhead{}&
 \colhead{} &
 \colhead{} &
 \colhead{} &
 \multicolumn{2}{c}{Car2}&
 \multicolumn{2}{c}{Car3}&
 \multicolumn{2}{c}{Car4}&
 \multicolumn{2}{c}{Car10}\\
\colhead{$\lambda$ (\AA)}&
 \colhead{Elem.} &
 \colhead{EP (eV)} &
 \colhead{$\log gf$} &
 \colhead{\feh} & \colhead{EW\tablenotemark{a}} &
 \colhead{\feh} & \colhead{EW\tablenotemark{a}} &
 \colhead{\feh} & \colhead{EW\tablenotemark{a}} &
 \colhead{\feh} & \colhead{EW\tablenotemark{a}} 
 }
\startdata

4924.770& 26.0& 2.279&$ -2.241$ & \ldots & \ldots & \ldots & \ldots & \ldots & \ldots & \ldots & \ldots
\\
4950.110& 26.0& 3.417&$ -1.670$ & \ldots & \ldots & \ldots & \ldots & \ldots & \ldots & \ldots & \ldots
\\
4973.100& 26.0& 3.960&$ -0.950$ & \ldots & \ldots &$-2.20\pm0.14$ &  88 & \ldots & \ldots & \ldots & \ldots
\\
5044.210& 26.0& 2.851&$ -2.038$ & \ldots & \ldots &$-2.15\pm0.13$ &  93 & \ldots & \ldots &$-2.10\pm0.12$ &  51
\\
5054.640& 26.0& 3.640&$ -1.921$ & \ldots & \ldots & \ldots & \ldots & \ldots & \ldots & \ldots & \ldots
\\
5090.770& 26.0& 4.256&$ -0.400$ & \ldots & \ldots & \ldots & \ldots & \ldots & \ldots & \ldots & \ldots
\\
5197.580& 26.1& 3.230&$ -2.348$ &$-1.55\pm0.30$ & 105 &$-1.85\pm0.38$ & 117 &$-1.49\pm0.42$ & 105 &$-2.00\pm0.21$ &  73
\\
5215.180& 26.0& 3.266&$ -0.871$ & \ldots & \ldots & \ldots & \ldots & \ldots & \ldots & \ldots & \ldots
\\
5217.390& 26.0& 3.211&$ -1.070$ & \ldots & \ldots & \ldots & \ldots & \ldots & \ldots & \ldots & \ldots
\\
5234.620& 26.1& 3.221&$ -2.279$ & \ldots & \ldots &$-1.95\pm0.34$ &  83 & \ldots & \ldots & \ldots & \ldots
\\
5242.490& 26.0& 3.634&$ -0.967$ & \ldots & \ldots &$-2.05\pm0.17$ &  98 & \ldots & \ldots & \ldots & \ldots
\\
5264.810& 26.1& 3.230&$ -3.133$ & \ldots & \ldots & \ldots & \ldots & \ldots & \ldots & \ldots & \ldots
\\
5284.110& 26.1& 2.891&$ -3.195$ & \ldots & \ldots & \ldots & \ldots &$-1.55\pm0.20$ &  65 & \ldots & \ldots
\\
5288.520& 26.0& 3.694&$ -1.508$ &$-1.65\pm0.13$ &  52 & \ldots & \ldots &$-1.70\pm0.24$ &  56 & \ldots & \ldots
\\
5307.360& 26.0& 1.608&$ -2.987$ & \ldots & \ldots & \ldots & \ldots & \ldots & \ldots & \ldots & \ldots
\\
5316.620& 26.1& 3.153&$ -2.014$ & \ldots & \ldots &$-2.05\pm0.20$ & 156 & \ldots & \ldots & \ldots & \ldots
\\

\enddata 
\tablenotetext{*}{This table is available entirety in a machine-readable form in the 
online journal. A portion is shown here for guidance regarding its form and content.}
\tablenotetext{a}{Equivalent widths are in m\AA.}
\end{deluxetable}

\begin{deluxetable}{lccccccc}
\tabletypesize{\small}
\tablewidth{0pt}
\tablecaption{Impact of uncertainties in effective temperature, surface gravity and microturbulence velocity on iron abundances for the representative star Car23.}
\label{tab_err}
\tablehead{
 \colhead{}&
 \multicolumn{2}{c}{$\Delta\Teff$ [K]}&
 \multicolumn{2}{c}{$\Delta\log g$ [$\log (\rm{cm\:s^{-2}})$]}&
 \multicolumn{2}{c}{$\Delta\xi$ [\kms]}\\
 \colhead{Ion}&
 \colhead{$-100$}&
 \colhead{$+100$}&
 \colhead{$-0.3$}&
 \colhead{$+0.3$}&
 \colhead{$-0.5$}&
 \colhead{$+0.5$}&
 \colhead{$<$$\sigma$$>$\tablenotemark{a}}
}
\startdata
 \feI  & $+0.10$ & $-0.11$ & $-0.01$ & $-0.00$ & $-0.10$ & $-0.01$ & $0.11$\\
 \feII & $+0.02$ & $+0.10$ & $-0.02$ & $+0.13$ & $-0.09$ & $-0.04$ & $0.12$\\
\enddata 
\tablenotetext{a}{Weighted standard deviation.}
\end{deluxetable}

\clearpage
\begin{figure}[!ht]
\begin{center}
\label{fig1}
\includegraphics[width=0.70\textwidth]{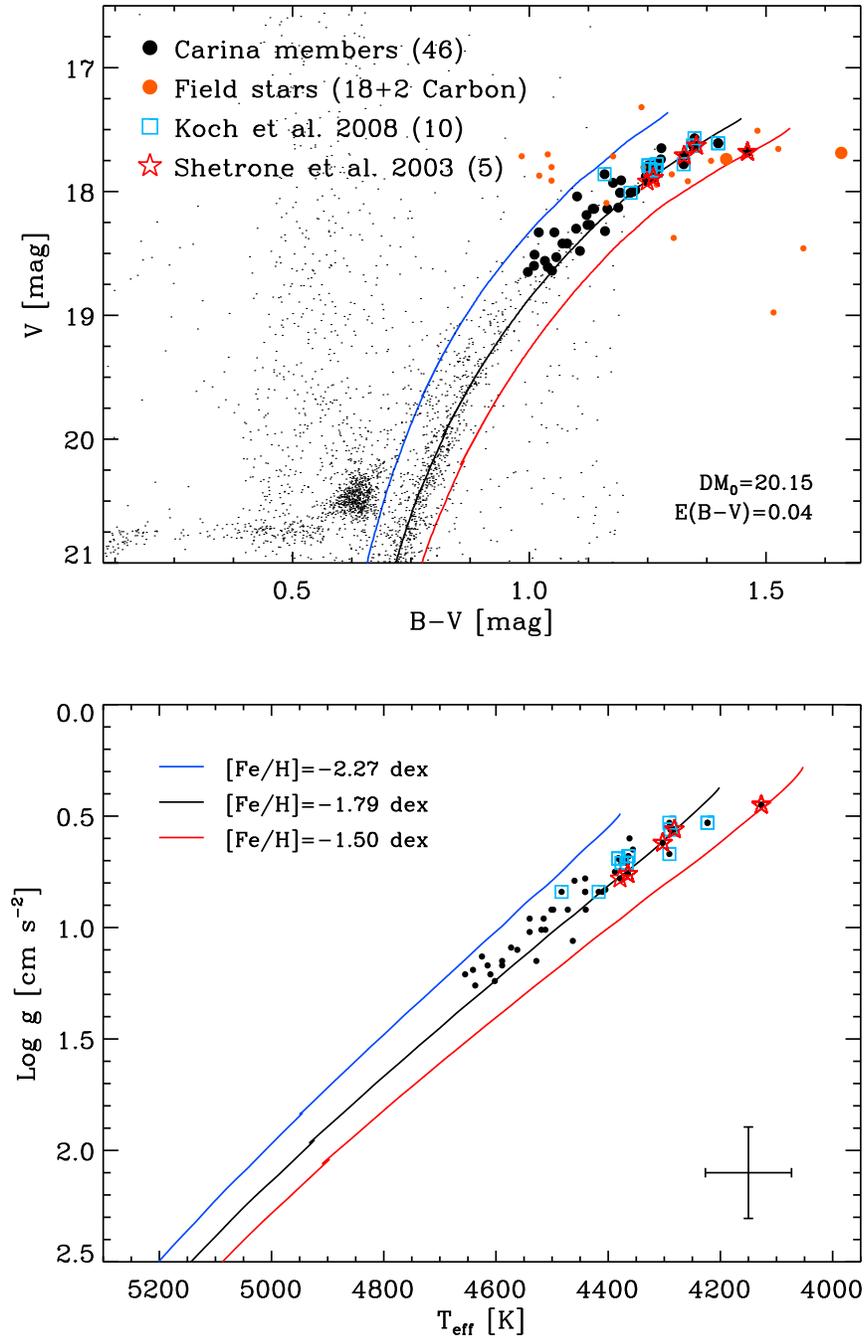}
\caption{Top: \V, \bmv\ CMD of the spectroscopic targets. The black dots mark
the UVES targets with a $RV$ between 180 and 260 \kms\ (Carina candidate
members), while the small orange dots show targets with $RV$ outside the
previous range (field stars). The two large orange dots refer to the two carbon
stars. The blue squares and red stars mark the targets in common with the sample
of \citet{koch08} and \citet{shetrone03}. The colored lines show three scaled
Solar isochrones \citep{pietr04,pietr06} at the same age ($t=12$~Gyr) with
$\feh=-1.50$ in red, $\feh =-1.79$ in black and $\feh= -2.27$ in blue. The
background gray dots are from the Carina photometric catalog \citep{bono10}.
Bottom: Same as the top, but the targets are plotted in the effective
temperature vs surface gravity plane. The error bars display typical
uncertainty on both effective temperature and surface gravity.
}  
\end{center}
\end{figure}

\clearpage
\begin{figure}[!ht]
\begin{center}
\label{fig2}
\includegraphics[width=0.70\textwidth]{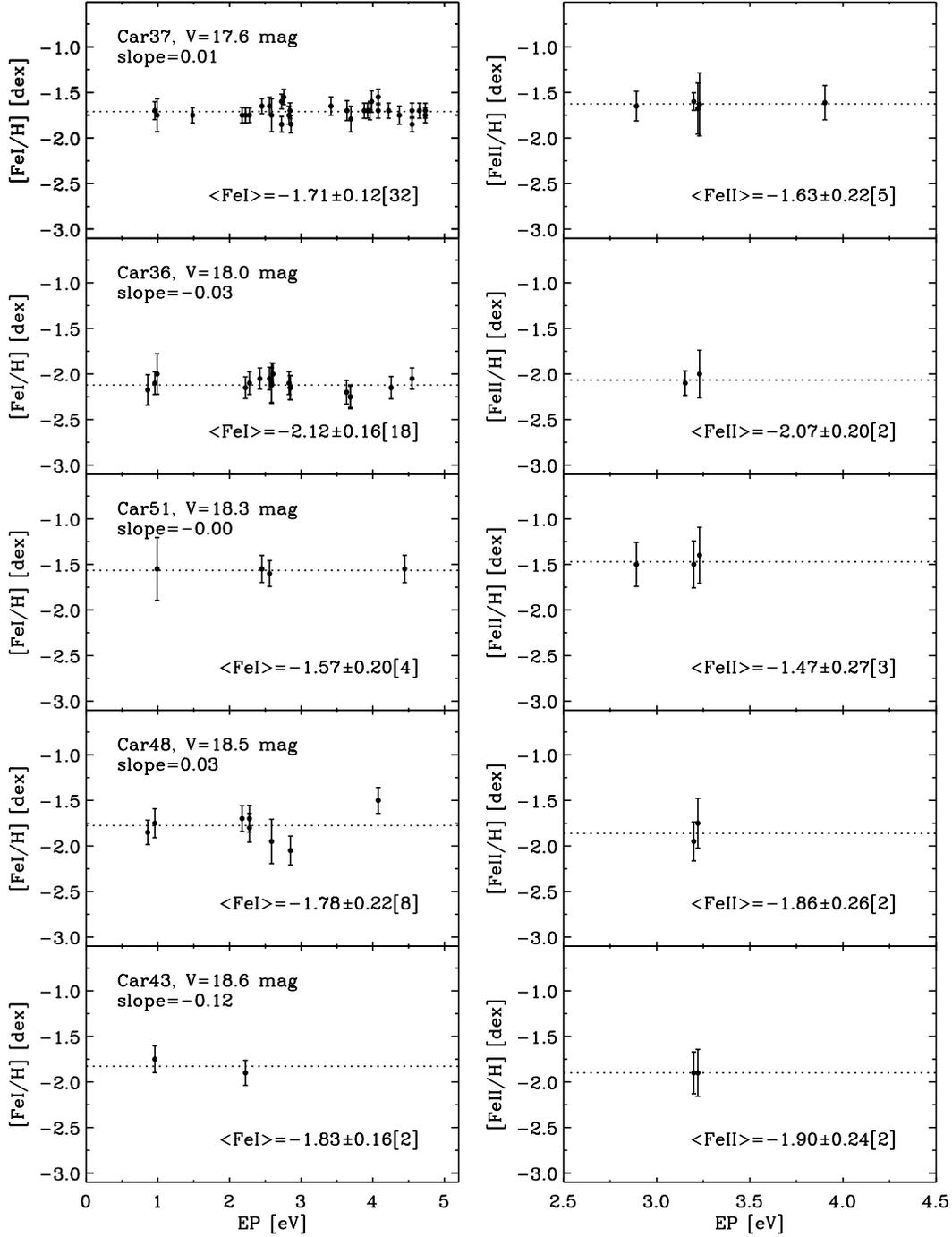}
\vspace*{1truecm}
\caption{Detailed iron abundances as a function of the Excitation Potential (EP)
for five stars covering the range in visual magnitude of the entire sample. From
top to bottom the panels show for each star the line-by-line analysis for the
\feI\ abundance (left panels) and for the \feII\ abundance (right panels). For
each ionization stage, the mean weighted value and the statistical error are
given and plotted as a dotted line. Numbers in square brackets are the number of
iron lines used to measure the abundance. 
}
\end{center}
\end{figure}

\begin{figure}[!ht]
\begin{center}
\label{fig3}
\includegraphics[width=0.70\textwidth]{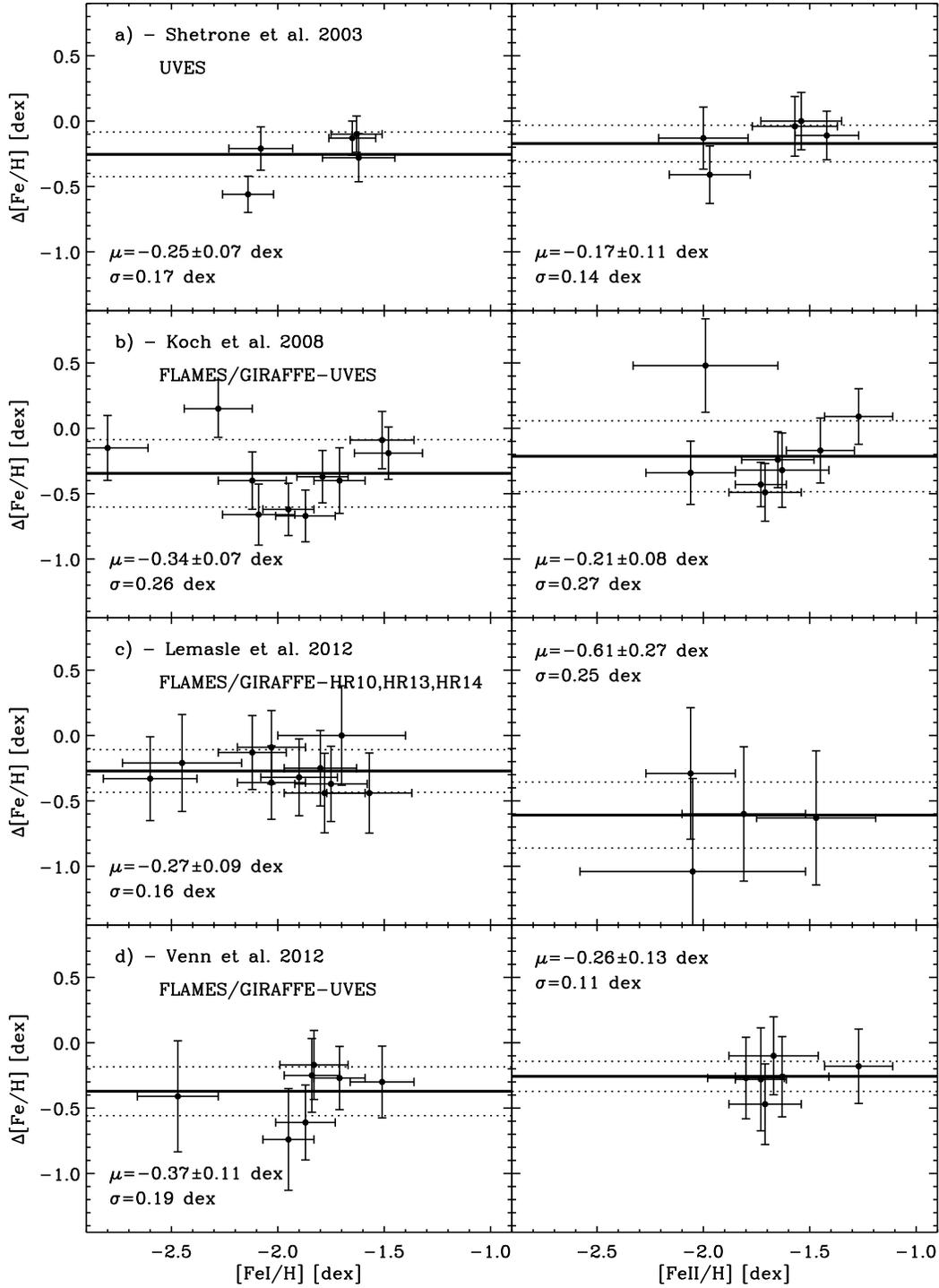}
\vspace*{1truecm}
\caption{
Panels a): Difference $\Delta\feh = \feh_{{\rm our}} - \feh_{\rm author}$ in
\feI\ (left) and in \feII\ (right) stellar abundances between our
measurements and similar measurements by \citet{shetrone03}. The
weighted mean $\mu$ (thick lines) and the weighted standard deviation
$\sigma$ are labeled together with the 1~$\sigma$ interval (dotted
lines). The error bars display individual uncertainties of the two sets
of abundances.
Panels b): Same as panels a), but the difference is with similar
measurements from \citet{koch08}.
Panels c): Same as panels a), but the difference is with similar
measurements from \citet{lemasle12}.
Panels d): Same as panels a), but the difference is with similar
measurements from \citet{venn12}.
} 
\end{center}
\end{figure}


\begin{figure}[!ht]
\begin{center}
\label{fig4}
\includegraphics[width=0.70\textwidth]{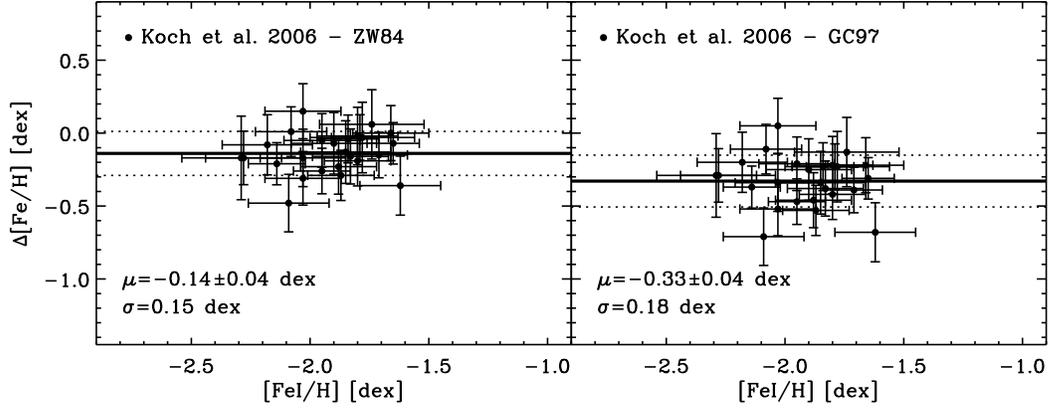}
\vspace*{-12truecm}
\caption{Difference in iron abundance (as in Fig.~3) for the 25 Carina stars in
common with \citet{koch06}. They estimated stellar iron abundances from the CaT
measurements based on medium-resolution spectra  in the ZW84 (left) and in the
CG97 (right) metallicity scale.
} 
\end{center}
\end{figure}

\begin{figure}[!ht]
\begin{center}
\label{fig5}
\includegraphics[width=0.70\textwidth]{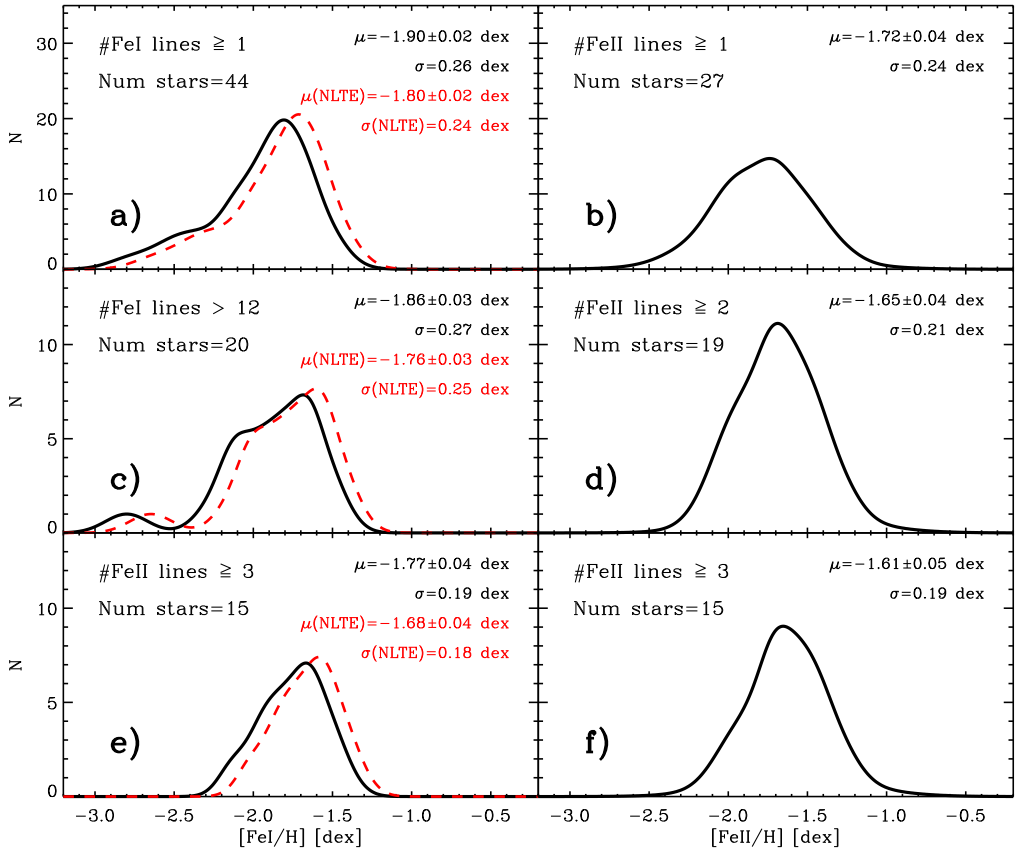}
\vspace*{-6truecm}
\caption{Panels a)--b): Metallicity distributions of Carina stars based on \feI\
(left) and \feII\ (right) lines. The weighted means and the weighted standard
deviations are labeled, together with the sample size.
The red dashed line shows the \feI\ metallicity distribution of the same stars,
but accounting for NLTE effects. The correction of individual abundances is
based on the linear fit between \feI\ and \feII\ abundances given in panel a) of
Fig.~6. The new $\mu$(NLTE) and $\sigma$(NLTE) values are labeled in red.
Panels c)--d): Same as the top, but for stars with more than twelve \feI\ or
with at least two \feII\ measurements.
Panels e)--f): Same as the top, but for stars with at least three \feII\
measurements.
}
\end{center}
\end{figure}

\begin{figure}[!ht]
\begin{center}
\label{fig6}
\includegraphics[width=0.70\textwidth]{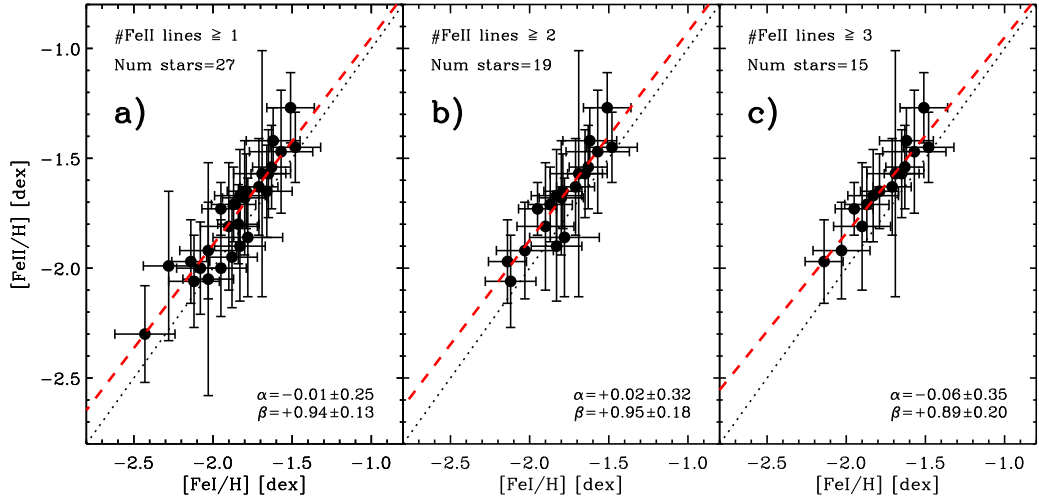}
\vspace*{-10truecm}
\caption{Panel a): Comparison between stellar iron abundances based on neutral
and on singly-ionized lines. The red dashed line shows the linear fit of the
data, used to correct the \feI\ metallicity distribution for NLTE effects. The
zero-point ($\alpha$) and the slope ($\beta$) of the linear fit are labeled.
Panels b) and c): Same as panel a), but for stars with at least two and at least
three \feII\ measurements, respectively.
}
\end{center}
\end{figure}

\begin{figure}[!ht]
\begin{center}
\label{fig7}
\includegraphics[width=0.70\textwidth]{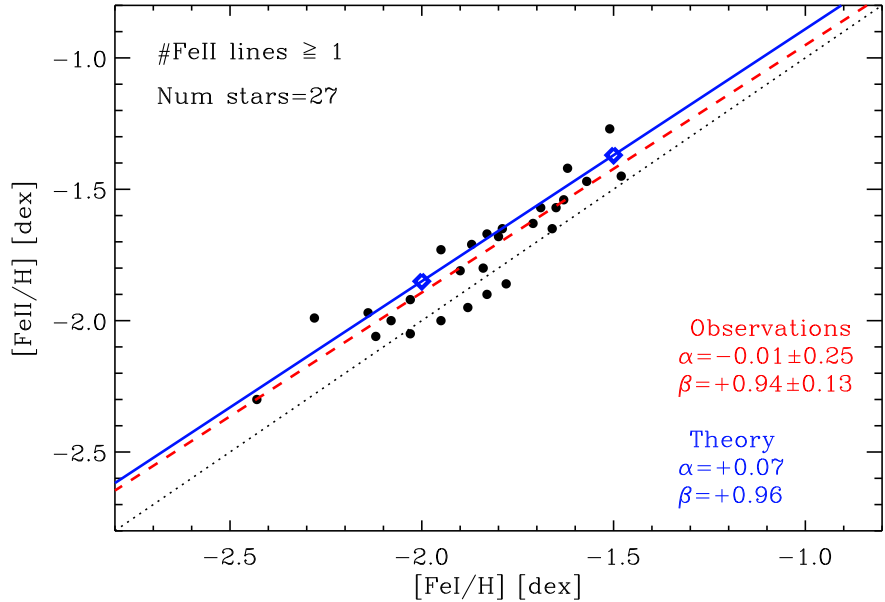}
\vspace*{-8.5truecm}
\caption{NLTE effects on ionization equilibrium. The symbols (black dots) are the
same as in panel a) of Fig.~6. The blue diamonds are the abundances given by
NLTE computations of \feIonII\ for two representative models of Carina RGs. We
adopted an effective temperature of 4500~K, a surface gravity of 1.0~dex and two
iron abundances $\feh=-2.0,-1.5$. The dashed red line shows the linear
regression on the data (the same as in panel a of Fig.~6), while the solid blue line
shows the linear fit on the two NLTE models. The zero-point ($\alpha$) and the
slope ($\beta$) of the linear fits are labeled.
}
\end{center}
\end{figure}

\end{document}